\documentclass[aps,showpacs]{revtex4}

\usepackage{amssymb,amsmath}
\usepackage[dvips]{graphicx}

\begin{document}

\title{\mbox{Phase synchronization between collective rhythms of globally coupled oscillator groups:} noiseless non-identical case}

\author{Yoji Kawamura}
\email{ykawamura@jamstec.go.jp}
\affiliation{Institute for Research on Earth Evolution,
Japan Agency for Marine-Earth Science and Technology, Yokohama 236-0001, Japan}

\author{Hiroya Nakao}
\affiliation{Department of Physics, Kyoto University, Kyoto 606-8502, Japan}
\affiliation{JST, CREST, Kyoto 606-8502, Japan}

\author{Kensuke Arai}
\affiliation{Brain Science Institute, RIKEN, Wako 351-0198, Japan}

\author{Hiroshi Kori}
\affiliation{Division of Advanced Sciences, Ochadai Academic Production, Ochanomizu University, Tokyo 112-8610, Japan}
\affiliation{PRESTO, Japan Science and Technology Agency, Kawaguchi 332-0012, Japan}

\author{Yoshiki Kuramoto}
\affiliation{Research Institute for Mathematical Sciences, Kyoto University, Kyoto 606-8502, Japan}
\affiliation{Institute for Integrated Cell-Material Sciences, Kyoto University, Kyoto 606-8501, Japan}

\date{July 26, 2010}


\begin{abstract}
Phase synchronization between collective oscillations
exhibited by two weakly interacting groups of non-identical phase oscillators
with internal and external global sinusoidal coupling of the groups is analyzed theoretically.
Coupled amplitude equations describing the collective oscillations of the oscillator groups
are obtained by using the Ott-Antonsen ansatz,
and then coupled phase equations for the collective oscillations
are derived by phase reduction of the amplitude equations.
The collective phase coupling function,
which determines the dynamics of macroscopic phase differences between the groups,
is calculated analytically.
It is demonstrated that the groups can exhibit effective anti-phase collective synchronization
even if the microscopic external coupling between individual oscillator pairs belonging to different groups is in-phase,
and similarly effective in-phase collective synchronization
in spite of microscopic anti-phase external coupling between the groups.
\end{abstract}

\maketitle


\section{Introduction} \label{sec:introduction}

Assemblies of coupled dynamical elements are abundant in Nature.
Macroscopic collective oscillations typically emerge in such systems
through mutual synchronization of the individual elements~\cite{
ref:winfree80,ref:kuramoto84,ref:pikovsky01,ref:strogatz03,
ref:manrubia04,ref:izhikevich07}.
Based on coupled phase oscillator models,
theoretical investigations of the origin and the nature
of collective oscillations have been carried out.
Globally coupled phase oscillators, a representative class of such models,
have been particularly well analyzed~\cite{ref:kuramoto75,ref:sakaguchi86,
ref:strogatz00,ref:balmforth00,ref:acebron05,
ref:boccaletti06,ref:arenas08,ref:dorogovtsev08}.
The appearance of an experimental system
of coupled electrochemical oscillators~\cite{ref:kiss02,ref:kiss05,ref:kiss07,ref:kiss08,ref:kori08}
has worked so powerfully in accelerating the study of collective dynamics of globally coupled oscillators.
Recently, Ott and Antonsen~\cite{ref:ott08,ref:ott09} proposed
a remarkable mathematical ansatz for the analytical treatment
of coupled phase oscillators in the continuum limit,
which is applicable to models with global sinusoidal coupling.
Since then, various applications~\cite{ref:abrams08,ref:childs08,ref:martens09,
ref:laing09a,ref:laing09b,ref:abdulrehem09,ref:marvel09a,ref:lee09,ref:pazo09,ref:nagai10}
and extensions~\cite{ref:pikovsky08,ref:marvel09b}
(see also Refs.~\cite{ref:watanabe93,ref:watanabe94})
of the Ott-Antonsen ansatz have been rapidly developed.

When two or more groups of dynamical elements
exhibiting collective oscillations interact with each other,
synchronization among those collective oscillations may naturally be expected.
In Refs.~\cite{ref:okuda91,ref:montbrio04,ref:barreto08,ref:sheeba08,ref:sheeba09},
two interacting groups of globally coupled phase oscillators have been studied
and mutual entrainment between the groups have been reported.
Description of the system in those works was essentially
based on microscopic phases of the individual oscillators,
and macroscopic properties of the collective synchronization
were also investigated on the microscopic footing.
However, it should be more convenient and beneficial
if one can describe the collective oscillations at the macroscopic level
using appropriate macroscopic variables in a closed form.

In this paper, using the collective phase of each oscillator group
as the macrovariable~\cite{ref:kawamura07,ref:kawamura08,ref:kawamura10,ref:kori09},
we formulate a theory of synchronization between two interacting groups
of globally coupled oscillators closed at the macroscopic level,
based on the Ott-Antonsen ansatz that gives a low-dimensional description
of phase oscillators with global sinusoidal coupling,
as well as on the standard phase reduction theory for limit-cycle oscillators.
We analytically derive the collective phase coupling function,
which determines the macroscopic dynamics of the interacting groups,
and illustrate several representative cases
of collective phase synchronization between the groups.
In particular, we reveal a counter-intuitive phenomenon in which
the macroscopic collective phase difference between two groups becomes anti-phase
in spite of microscopic in-phase external coupling between individual pairs of oscillators between the groups,
and also the opposite phenomenon, namely, in-phase synchronization between oscillator groups
with microscopic anti-phase external coupling.

In Ref.~\cite{ref:kawamura10}, we considered a similar problem, namely,
collective phase synchronization between two groups of globally coupled phase oscillators.
The crucial difference is that we treat
{\it deterministic noiseless non-identical phase oscillators} in the present paper,
whereas we analyzed {\it stochastic noisy identical phase oscillators} in Ref.~\cite{ref:kawamura10}.
Though these two cases may look similar,
they are physically different systems and require distinct mathematical approaches;
here we rely on the Ott-Antonsen ansatz,
whereas we used center-manifold reduction as well as phase reduction to nonlinear Fokker-Planck equations in Ref.~\cite{ref:kawamura10}.
As we will show, we still find very similar transitions between effective in-phase and anti-phase synchronization in both cases.
This implies that the two systems have similar effective low-dimensional dynamics,
despite the fact that the two models are originally defined in completely different high-dimensional phase spaces.

The organization of this paper is as follows.
In Sec.~\ref{sec:simulation},
we introduce a model of weakly interacting groups of globally coupled phase oscillators
and illustrate both effective anti-phase and in-phase synchronization of collective oscillations
by numerical simulations.
In Sec.~\ref{sec:theory}, we develop a theory based on the macroscopic phase description of collective oscillations
that clarifies whether the phase coupling between collective oscillations is effectively in-phase or anti-phase.
In Sec.~\ref{sec:types}, we illustrate several representative cases of the collective phase coupling function
obtained in Sec.~\ref{sec:theory} and reexamine the numerical results in Sec.~\ref{sec:simulation}.
Concluding remarks will be given in the final section.

\section{Collective phase synchronization of oscillator groups} \label{sec:simulation}

We consider two interacting groups of globally coupled non-identical phase oscillators
described by the following model:
\begin{equation}
  \dot{\phi}_j^{(\sigma)}(t) = \omega_j - \frac{K}{N}
  \sum_{k=1}^N \sin\left( \phi_j^{(\sigma)} - \phi_k^{(\sigma)} + \alpha \right)
  - \frac{\epsilon J}{N}
  \sum_{k=1}^N \sin\left( \phi_j^{(\sigma)} - \phi_k^{(\tau)} + \beta \right),
  \label{eq:model}
\end{equation}
for $j = 1, \cdots, N$ and $(\sigma, \tau) = (1, 2)$ or $(2, 1)$,
where $\phi_j^{(\sigma)}(t)$ is the phase of the $j$-th oscillator
in the $\sigma$-th group consisting of $N$ oscillators.
The second term on the right-hand side represents internal coupling between oscillators within the same group,
and the last term gives external coupling between oscillators that belong to different groups.
This phase model can be derived from two interacting groups
of globally coupled limit-cycle oscillators near the Hopf bifurcation point
using the phase reduction method~\cite{ref:kuramoto84} under weak coupling conditions.
Phase oscillators with global sinusoidal coupling as given in Eq.~(\ref{eq:model}) can be experimentally realized
in electrochemical oscillator systems~\cite{ref:kiss02,ref:kiss05,ref:kiss07,ref:kiss08,ref:kori08}.

The internal coupling is specified by the parameters $K > 0$ and $|\alpha| < \pi / 2$,
where $K$ determines the coupling intensity and $\alpha$ gives the coupling phase shift.
These parameter values correspond to in-phase (attractive) internal coupling.
Similarly, the external coupling is specified by the parameters $J > 0$ and $|\beta| \leq \pi$.
The external coupling can be either in-phase (attractive) ($|\beta| < \pi/2$)
or anti-phase (repulsive) ($|\beta| > \pi/2$)~\cite{ref:kuramoto84}.
The characteristic magnitude of the weak external coupling
is given by a small parameter $\epsilon \geq 0$.

The natural frequency $\omega_j$ is assumed to be drawn from the Lorentzian distribution
with central value $\omega_0$ and dispersion $\gamma$,
\begin{equation}
  g\left( \omega \right)
  = \frac{\gamma}{\pi} \frac{1}{\left( \omega - \omega_0 \right)^2 + \gamma^2}.
  \label{eq:lorentzian}
\end{equation}
We define a parameter
\begin{equation}
  \eta = \frac{\gamma}{K \cos \alpha},
  \label{eq:ratio}
\end{equation}
which is the ratio of the frequency dispersion $\gamma$
to the attracting component $K \cos \alpha$ of the internal coupling function.
When the external coupling is absent, i.e., $\epsilon = 0$,
Eq.~(\ref{eq:model}) describes two independent oscillator groups ($\sigma=1,2$),
each of which exhibits collective oscillations under the condition
$0 \leq \eta < 1/2$~\cite{ref:sakaguchi86}.

Introducing a complex order parameter $A^{(\sigma)}(t)$
with modulus $R^{(\sigma)}(t)$ and phase $\Theta^{(\sigma)}(t)$ through
\begin{equation}
  A^{(\sigma)}\left( t \right)
  = R^{(\sigma)}\left( t \right) e^{i \Theta^{(\sigma)}\left( t \right)}
  = \frac{1}{N} \sum_{k=1}^N e^{i \phi_k^{(\sigma)}\left( t \right)}
  \label{eq:order}
\end{equation}
for each group, we can rewrite Eq.~(\ref{eq:model}) as
\begin{equation}
  \dot{\phi}_j^{(\sigma)} = \omega_j - K R^{(\sigma)}
  \sin\left( \phi_j^{(\sigma)} - \Theta^{(\sigma)} + \alpha \right)
  - \epsilon J R^{(\tau)}
  \sin\left( \phi_j^{(\sigma)} - \Theta^{(\tau)} + \beta \right).
  \label{eq:simulation}
\end{equation}
The order parameter $R^{(\sigma)}$ quantifies the degree of synchronization in each group ($0 \leq R^{(\sigma)} \leq 1$),
and $\Theta^{(\sigma)}$ gives the macroscopic collective phase of the group.

Focusing on weakly coupled collective oscillations,
we carried out numerical simulations of Eq.~(\ref{eq:simulation})
with Eq.~(\ref{eq:order}) under the following conditions:
Without loss of generality, we can assume $K = J = 1$ and $\omega_0 = 0$.
The external coupling was assumed to be much weaker than the internal coupling, $\epsilon = 0.01$.
We set the frequency dispersion $\gamma = (K \cos \alpha) / 4$
and the internal coupling phase shift $\alpha = 3 \pi / 8$,
so that the parameter $\eta$ was given by $\eta = 1 / 4$.
The number of oscillators in each group was $N = 4096$,
which was sufficiently large to observe clear collective oscillations.

Figure~\ref{fig:1}(a) shows typical evolution of the collective phase difference,
$| \Theta^{(1)} - \Theta^{(2)} |$,
with in-phase condition for microscopic external coupling, $\beta = 3 \pi / 8$.
Two groups of phase oscillators exhibiting collective oscillations
were separately prepared with their collective phases being almost equal,
and these states were used as the initial condition.
Despite the in-phase external coupling condition for individual pairs of oscillators in different groups,
the collective phase difference became anti-phase ($| \Theta^{(1)} - \Theta^{(2)} | = \pi$) after some time.
Thus, Fig.~\ref{fig:1}(a) implies that effective anti-phase coupling between collective oscillations is realized.
In contrast, Fig.~\ref{fig:1}(b) shows effective in-phase synchronization between collective oscillations
($| \Theta^{(1)} - \Theta^{(2)} | = 0$) with microscopic anti-phase external coupling, $\beta = -5 \pi / 8$.

Snapshots of the phase oscillators after the collective phase difference has reached the asymptotic value in Fig.~\ref{fig:1}
are displayed in Fig.~\ref{fig:2}.
The oscillators are sorted in increasing order of their natural frequencies.
The coherent segment represents phase-locked oscillators within each group
and scattered points correspond to drifting oscillators.
The coherent, phase-locked segment of each group is not centered about the middle oscillator
because the internal coupling phase shift $\alpha$ is non-zero.
In Fig.~\ref{fig:2}(a), the two distributions of the oscillators are shifted by $\pi$,
indicating anti-phase synchronization between the groups.
In contrast, the two distributions almost overlap in Fig.~\ref{fig:2}(b),
i.e., they are in-phase synchronized.
Note that drifting oscillators from different groups do not synchronize with each other,
in other words, the collective phase synchronization between the groups is not
due to complete synchronization of individual oscillators at the microscopic level.

\section{Derivation of the collective phase coupling function} \label{sec:theory}

We now derive collective phase equations describing the interacting oscillator groups
via the amplitude equations obtained by using the Ott-Antonsen ansatz~\cite{ref:ott08,ref:ott09}.
We analytically determine the collective phase coupling function and its type,
specifically, whether it is in-phase or anti-phase.

Using the order parameters defined in Eq.~(\ref{eq:order}),
we can rewrite Eq.~(\ref{eq:model}) as
\begin{equation}
  \dot{\phi}_j^{(\sigma)} = \omega_j - \frac{K}{2 i}
  \Bigl( \bar{A}^{(\sigma)} \, e^{i \phi_j^{(\sigma)}} \, e^{i \alpha}
  - A^{(\sigma)} \, e^{-i \phi_j^{(\sigma)}} \, e^{-i \alpha} \Bigr)
  - \frac{\epsilon J}{2 i}
  \Bigl( \bar{A}^{(\tau)} \, e^{i \phi_j^{(\sigma)}} \, e^{i \beta}
  - A^{(\tau)} \, e^{-i \phi_j^{(\sigma)}} \, e^{-i \beta} \Bigr),
\end{equation}
where $\bar{A}^{(\sigma)}$ is the complex conjugate of $A^{(\sigma)}$.
In the continuum limit, $N \to \infty$,
we can obtain the following continuity equation for each $\sigma$~\cite{ref:ott08,ref:ott09}:
\begin{align}
  \frac{\partial}{\partial t} f^{(\sigma)}\left( \phi, \omega, t \right)
  + \frac{\partial}{\partial \phi} \Biggl[ \Biggl\{ \omega
    &- \frac{K}{2 i}
    \Bigl( \bar{A}^{(\sigma)} \, e^{i \phi} \, e^{i \alpha}
    - A^{(\sigma)} \, e^{-i \phi} \, e^{-i \alpha} \Bigr) \nonumber \\
    &- \frac{\epsilon J}{2 i}
    \Bigl( \bar{A}^{(\tau)} \, e^{i \phi} \, e^{i \beta}
    - A^{(\tau)} \, e^{-i \phi} \, e^{-i \beta} \Bigr)
    \Biggr\} f^{(\sigma)}\left( \phi, \omega, t \right) \Biggr] = 0,
  \label{eq:continuity}
\end{align}
which describes dynamics of the probability density function 
$f^{(\sigma)}(\phi, \omega, t)$ of the phase and the frequency.
Here, $f^{(\sigma)}(\phi, \omega, t)$ satisfies normalization conditions
\begin{equation}
  \int_0^{2 \pi} d\phi \, f^{(\sigma)}\left( \phi, \omega, t \right) = g\left( \omega \right), \qquad
  \int_0^{2 \pi} d\phi \int_{-\infty}^{\infty} d\omega \, f^{(\sigma)}\left( \phi, \omega, t \right) = 1,
\end{equation}
and the complex order parameter $A^{(\sigma)}$ is now defined by
\begin{equation}
  A^{(\sigma)}\left( t \right)
  = \int_0^{2 \pi} d\phi \int_{-\infty}^{\infty} d\omega \, e^{i \phi} f^{(\sigma)}\left( \phi, \omega, t \right)
\end{equation}
for $\sigma = 1, 2$.

We now apply the Ott-Antonsen ansatz~\cite{ref:ott08,ref:ott09}, 
\begin{equation}
  f^{(\sigma)}\left( \phi, \omega, t \right) = \frac{g\left( \omega \right)}{2 \pi}
  \left[ 1 + \sum_{n=1}^{\infty} \Biggl\{ \Bigl( a^{(\sigma)}\left(\omega, t \right) \Bigr)^n e^{i n \phi}
    + \Bigl( \bar{a}^{(\sigma)}\left(\omega, t \right) \Bigr)^n e^{-i n \phi} \Biggr\} \right],
\end{equation}
to the continuity equation~(\ref{eq:continuity}),
which replaces all the Fourier coefficients of $f^{(\sigma)}(\phi, \omega, t)$
by integer powers of the complex variable $a^{(\sigma)}(\omega, t)$.
This ansatz leads to a two-dimensional representation of the infinite-dimensional partial differential equation
when the frequency distribution $g(\omega)$ is Lorentzian.
It has been shown that the above restricted functional form of $f^{(\sigma)}(\phi, \omega, t)$
yields asymptotically correct dynamics of the complex order parameter $A^{(\sigma)}(t)$~\cite{ref:ott09}.

By substituting this expression into Eq.~(\ref{eq:continuity}),
we can derive the following equation for the complex dynamical variable $a^{(\sigma)}(\omega, t)$:
\begin{equation}
  \frac{\partial}{\partial t} a^{(\sigma)} + i \omega a^{(\sigma)}
  + \frac{K}{2}
  \left[ A^{(\sigma)} \, \left( a^{(\sigma)} \right)^2 \, e^{-i \alpha}
    - \bar{A}^{(\sigma)} \, e^{i \alpha} \right]
  + \frac{\epsilon J}{2}
  \left[ A^{(\tau)} \, \left( a^{(\sigma)} \right)^2 \, e^{-i \beta}
    - \bar{A}^{(\tau)} \, e^{i \beta} \right] = 0,
\end{equation}
where $(\sigma, \tau) = (1, 2)$ or $(2, 1)$.
Moreover, in the case of the Lorentzian frequency distribution, Eq.~(\ref{eq:lorentzian}),
the complex order parameter $A^{(\sigma)}(t)$ can simply be expressed by $a^{(\sigma)}(\omega, t)$ as
\begin{equation}
  A^{(\sigma)}\left( t \right)
  = \int_0^{2 \pi} d\phi \int_{-\infty}^{\infty} d\omega \,
  e^{i \phi} f^{(\sigma)}\left( \phi, \omega, t \right)
  = \int_{-\infty}^{\infty} d\omega \,
  \bar{a}^{(\sigma)}\left( \omega, t \right) g\left( \omega \right)
  = \bar{a}^{(\sigma)}\left( \omega = \omega_0 - i \gamma, \, t \right)
\end{equation}
by performing a complex contour integral in the lower-half complex plane of $\omega$,
where $\omega = \omega_{0} - i \gamma$ gives the pole of the integrand with the Lorentzian $g(\omega)$~\cite{ref:ott08,ref:ott09}.
Therefore, we arrive at the following coupled amplitude equation
for the complex order parameter $A^{(\sigma)}(t)$ in a closed form:
\begin{equation}
  \dot{A}^{(\sigma)}
  = \left( \mu + i \Omega_{\rm c} \right) A^{(\sigma)}
  - g \left| A^{(\sigma)} \right|^2 A^{(\sigma)}
  + \epsilon \left[ \bar{d} \, A^{(\tau)} - d
    \left(A^{(\sigma)}\right)^2 \bar{A}^{(\tau)} \right],
  \label{eq:amplitude}
\end{equation}
for $(\sigma, \tau) = (1,2)$ or $(2,1)$, where the parameters are defined by
\begin{equation}
  \mu = \frac{K \cos \alpha}{2} - \gamma, \quad
  \Omega_{\rm c} = \omega_0 - \frac{K \sin \alpha}{2}, \quad
  g = \frac{K}{2} e^{i \alpha}, \quad
  d = \frac{J}{2} e^{i \beta}.
  \label{eq:parameter}
\end{equation}
Note that Eq.~(\ref{eq:amplitude}) describes two coupled Stuart-Landau oscillators,
each of which (i.e., $\dot{A} = ( \mu + i \Omega_{\rm c} ) A - g |A|^2 A$)
represents collective oscillations of the respective oscillator group.
Also, note that it is valid for the whole parameter region of the system,
not only near the synchronization transition points of each oscillator group, $\eta = 1/2$ ($\mu = 0$).
This is in sharp contrast to the conventional center-manifold reduction method~\cite{ref:kuramoto84}
that leads to similar coupled amplitude equations,
but which is valid only near the bifurcation points.

As mentioned above, each oscillator group exhibits collective oscillations when $\eta < 1/2$ ($\mu > 0$).
Correspondingly, in the absence of the external coupling, $\epsilon = 0$, 
Eq.~(\ref{eq:amplitude}) has a circular limit-cycle solution $A_0(\Theta)$ on the complex plane,
whose analytical expression can explicitly be given by
\begin{equation}
  A_0(\Theta) = \sqrt{\frac{\mu}{{\rm Re}\, g}} e^{i \Theta}, \qquad
  R_0 = \left| A_0 \right| = \sqrt{1 - 2\eta}, \qquad
  \dot{\Theta} = \Omega
  = \Omega_{\rm c} - \mu \frac{{\rm Im}\, g}{{\rm Re}\, g}
  = \omega_0 - K \sin \alpha + \gamma \tan \alpha,
  \label{eq:solution}
\end{equation}
where $R_0$ and $\Omega$ represent the amplitude and the frequency
of the collective oscillation in the model~(\ref{eq:model}) with $\epsilon = 0$, respectively (see Appendix.~\ref{sec:A}).
The right Floquet eigenvector of the limit-cycle solution $A_0(\Theta)$ associated with the zero eigenvalue 
is given by $U_0(\Theta) = dA_0(\Theta)/d\Theta$, namely,
\begin{equation}
  U_0(\Theta) = i \sqrt{\frac{\mu}{{\rm Re}\, g}} e^{i \Theta},
\end{equation}
and the corresponding left zero Floquet eigenvector at each $\Theta$ can be taken as
\begin{equation}
  U_0^\ast(\Theta) = i \sqrt{\frac{{\rm Re}\, g}{\mu}} \frac{g}{{\rm Re}\, g} e^{i \Theta}.
  \label{eq:eigenvector}
\end{equation}
Taken together, they satisfy the normalization condition
\begin{equation}
  {\rm Re}\, \Bigl[ \bar{U}_0^\ast\left( \Theta \right) U_0\left( \Theta \right) \Bigr] = 1.
\end{equation}
Though the above quantities are expressed by complex numbers
for the sake of convenience in analytical calculations performed below,
they are exactly the same as the known results for the Stuart-Landau oscillator~\cite{ref:kuramoto84}.

Now let us introduce weak external coupling,
i.e., we assume that $\epsilon$ takes small positive values
and treat the last term of the amplitude equation~(\ref{eq:amplitude}) as a perturbation.
Using the phase reduction method~\cite{ref:kuramoto84},
we can obtain the collective phase dynamics of the amplitude equation~(\ref{eq:amplitude})
by projecting it onto the unperturbed limit-cycle orbit as
\begin{align}
  \dot{\Theta}^{(\sigma)}
  &= {\rm Re} \left[ \bar{U}_0^\ast\left( \Theta^{(\sigma)} \right) \dot{A}^{(\sigma)} \right] \nonumber \\
  &\simeq \Omega + \epsilon {\rm Re} \left[ \bar{U}_0^\ast\left( \Theta^{(\sigma)} \right)
    \left\{ \bar{d} \, A_0\left( \Theta^{(\tau)} \right)
    - d \left( A_0\left( \Theta^{(\sigma)} \right) \right)^2 \bar{A_0}\left( \Theta^{(\tau)} \right) \right\} \right],
\end{align}
where we approximated $A^{(\sigma)}$ by the unperturbed solution $A_0(\Theta^{(\sigma)})$
and used ${\rm Re} [ \bar{U}_0^\ast(\Theta) \dot{A_0}(\Theta) ] = \Omega$.
We have thus obtained the following coupled collective phase equation from Eq.~(\ref{eq:amplitude}):
\begin{equation}
  \dot{\Theta}^{(\sigma)}
  = \Omega + \epsilon \, \digamma\left( \Theta^{(\sigma)} - \Theta^{(\tau)} \right),
  \label{eq:phase}
\end{equation}
for $(\sigma, \tau)=(1,2)$ or $(2,1)$, where the collective phase coupling function is given by
\begin{equation}
  \digamma\left( \Theta^{(\sigma)} - \Theta^{(\tau)} \right)
  = {\rm Re}\, \left[ \bar{U}_0^\ast\left( \Theta^{(\sigma)} \right)
    \left\{ \bar{d} \, A_0\left( \Theta^{(\tau)} \right)
    - d \left( A_0\left( \Theta^{(\sigma)} \right) \right)^2 \bar{A_0}\left( \Theta^{(\tau)} \right) \right\} \right].
  \label{eq:digamma}
\end{equation}
Similarly, we can also derive the collective phase sensitivity function~\cite{ref:kawamura08} (see Appendix.~\ref{sec:B}).

By inserting the expressions of Eqs.~(\ref{eq:parameter}), (\ref{eq:solution}), and (\ref{eq:eigenvector})
into Eq.~(\ref{eq:digamma}), the collective phase coupling function $\digamma(\Theta)$ is obtained in the sinusoidal form,
\begin{equation}
  \digamma\left( \Theta \right) = -\rho \sin\left( \Theta + \delta \right),
  \label{eq:digamma2}
\end{equation}
where the parameters $\rho$ and $\delta$ are respectively the modulus and the argument of a complex number given by
\begin{equation}
  \rho e^{i \delta} = J \Bigl[ \bigl\{
    \left( 1 - \eta \right) \cos \beta - \eta \tan \alpha \sin \beta \bigr\}
    + i \bigl\{ \left( 1 - \eta \right) \sin \beta + \eta \tan \alpha \cos \beta \bigr\} \Bigr].
  \label{eq:type}
\end{equation}
This formula is the main result of the present paper.
It determines the collective phase coupling function,
Eq.~(\ref{eq:digamma2}) in the collective phase equation~(\ref{eq:phase}),
which is derived from Eq.~(\ref{eq:model}) via the complex amplitude equation~(\ref{eq:amplitude}).
The type of coupling is found from the following quantity:
\begin{equation}
  \rho \cos \delta
  = J \bigl\{ \left( 1 - \eta \right) \cos \beta - \eta \tan \alpha \sin \beta \bigr\},
  \label{eq:type2}
\end{equation}
where $\rho \cos \delta > 0$ represents the in-phase coupling and $\rho \cos \delta < 0$ gives the anti-phase coupling.
Reflecting the symmetry of Eq.~(\ref{eq:model}),
Eq.~(\ref{eq:type2}) is symmetric about the origin in the $\alpha$-$\beta$ plane.

\section{Representative cases of collective phase coupling functions} \label{sec:types}

We here illustrate five representative cases of the collective phase coupling function obtained in Sec.~\ref{sec:theory},
which correspond to several special sets of the parameters, i.e.,
the ratio $\eta$ given by Eq.~(\ref{eq:ratio}), the phase shift $\alpha$ of the internal coupling function,
and the phase shift $\beta$ of the external coupling function.
We then reexamine the results of our numerical simulation in Sec.~\ref{sec:simulation}.

(i) The first case is $\eta = 0$, which implies that all oscillators are identical, i.e., $\gamma = 0$.
In this case, the oscillators in the same group become completely phase synchronized due to the in-phase internal coupling,
so that the maximum amplitude of collective oscillations is realized, namely, $R_0 = 1$.
Inserting $\eta = 0$ into Eq.~(\ref{eq:type}), we obtain the following result:
\begin{equation}
  \eta = 0, \qquad
  \rho \, e^{i \delta} = J e^{i \beta},
\end{equation}
which says that the parameters of the collective phase coupling function
are identical to those of the microscopic external phase coupling function,
so that the types of the effective coupling between the groups
and the external coupling between individual oscillators coincide.
The same result for the completely phase synchronized case has been obtained
in different ways~\cite{ref:kawamura08,ref:kawamura10,ref:kori09}.
Note that the above results are independent of the value of the internal coupling phase shift $\alpha$,
so that $\alpha$ does not affect the collective phase coupling function at all.

(ii) The second case is the limit $\eta \to 1/2$,
which indicates that each oscillator group is exactly at the onset of collective oscillations, i.e., $R_{0} \to 0$.
Inserting $\eta = 1/2$ into Eq.~(\ref{eq:type}), we obtain the following result:
\begin{equation}
  \eta \to \frac{1}{2}, \qquad
  \rho \, e^{i \delta} = \frac{J}{2 \cos \alpha} e^{i \left( \alpha + \beta \right)},
\end{equation}
which yields the real part $\rho\cos\delta = (J/2) \{ \cos\beta - \tan\alpha \sin\beta \}$.
Thus, the microscopic internal coupling parameter $\alpha$ most significantly affects
the parameters of the collective phase coupling function, in contrast to the case~(i).
Depending on the values of $\alpha$ and $\beta$,
the types of the effective coupling between the groups
and the external coupling between individual oscillators can be either the same or opposite.

(iii) The third case is $\alpha = 0$,
which gives an antisymmetric (odd) internal coupling function between individual oscillators.
In this case, $\eta = \gamma / K$ and $R_0 = \sqrt{1 - 2 \eta}$.
Inserting $\alpha = 0$ into Eq.~(\ref{eq:type}), we obtain the following result:
\begin{equation}
  \alpha = 0, \qquad
  \rho \, e^{i \delta} = (1 - \eta) J e^{i \beta}.
\end{equation}
Thus, the type of the collective phase coupling function
is solely determined by the microscopic external coupling phase shift $\beta$.
Further, the collective and microscopic external coupling functions are of the same type.
Similar scenarios have been encountered in different models~\cite{ref:kawamura08,ref:kawamura10,ref:kori09}.

(iv) The fourth cases correspond to special values of the microscopic external coupling phase shift $\beta$,
which give symmetric (even) or antisymmetric (odd) external coupling functions.
Inserting $\beta = 0$ (in-phase), $\pm\pi$ (anti-phase), and $\pm\pi/2$ (marginal) into Eq.~(\ref{eq:type}),
we obtain the following results:
\begin{equation}
  \beta = 0, \qquad
  \rho e^{i \delta} = J \left[ \left( 1 - \eta \right) + i \eta \tan \alpha \right],
\end{equation}
\begin{equation}
  \beta = \pm\pi, \qquad
  \rho e^{i \delta} = J \left[ -\left( 1 - \eta \right) - i \eta \tan \alpha \right],
\end{equation}
\begin{equation}
  \beta = \pm\frac{\pi}{2}, \qquad
  \rho e^{i \delta} = J \left[ \mp \eta \tan \alpha \pm i \left( 1 - \eta \right) \right].
\end{equation}
For antisymmetric (odd) microscopic external coupling functions, i.e., for $\beta = 0$ and $\pm\pi$,
the type of the collective phase coupling is not affected
by the internal coupling phase shift $\alpha$,
because $\alpha$ does not appear in the real part $\rho \cos \delta$.
In contrast, for the symmetric (even) microscopic external coupling, i.e., $\beta = \pm\pi/2$,
the type of the collective phase coupling function is solely determined by the internal coupling parameter $\alpha$.
The types of effective coupling between groups and the external coupling between individual oscillators coincide
when $\beta = 0, \pm \pi$ and also when $\beta = \pm\pi/2$ and $\alpha = 0$.

(v) The fifth case is $\beta = \alpha$, namely,
the case that the external and the internal couplings have the same coupling phase shift.
Since we assume $| \alpha | < \pi / 2$, the value of $\beta$ should also be in this range.
Inserting $\beta = \alpha$ into Eq.~(\ref{eq:type}), we obtain the following result:
\begin{equation}
  \beta = \alpha, \qquad
  \rho \, e^{i \delta} = J \left[ \left\{ \cos \alpha - \frac{\eta}{\cos \alpha} \right\} + i \sin \alpha \right].
\end{equation}
In this case, $\rho \cos \delta$ depends on the internal coupling phase shift $\alpha$ and on the ratio $\eta$.
Therefore, effective anti-phase collective coupling can be realized when $\cos^2 \alpha < \eta$,
in spite of the microscopic in-phase external coupling ($|\alpha| < \pi / 2$).
When $\eta \to 1/2$, the effective anti-phase collective coupling is realized for $\cos^2 \alpha < 1/2$,
i.e., $\pi/4 < |\alpha| < \pi/2$.

\begin{figure}
  \begin{center}
    \includegraphics[width=0.45\hsize,clip]{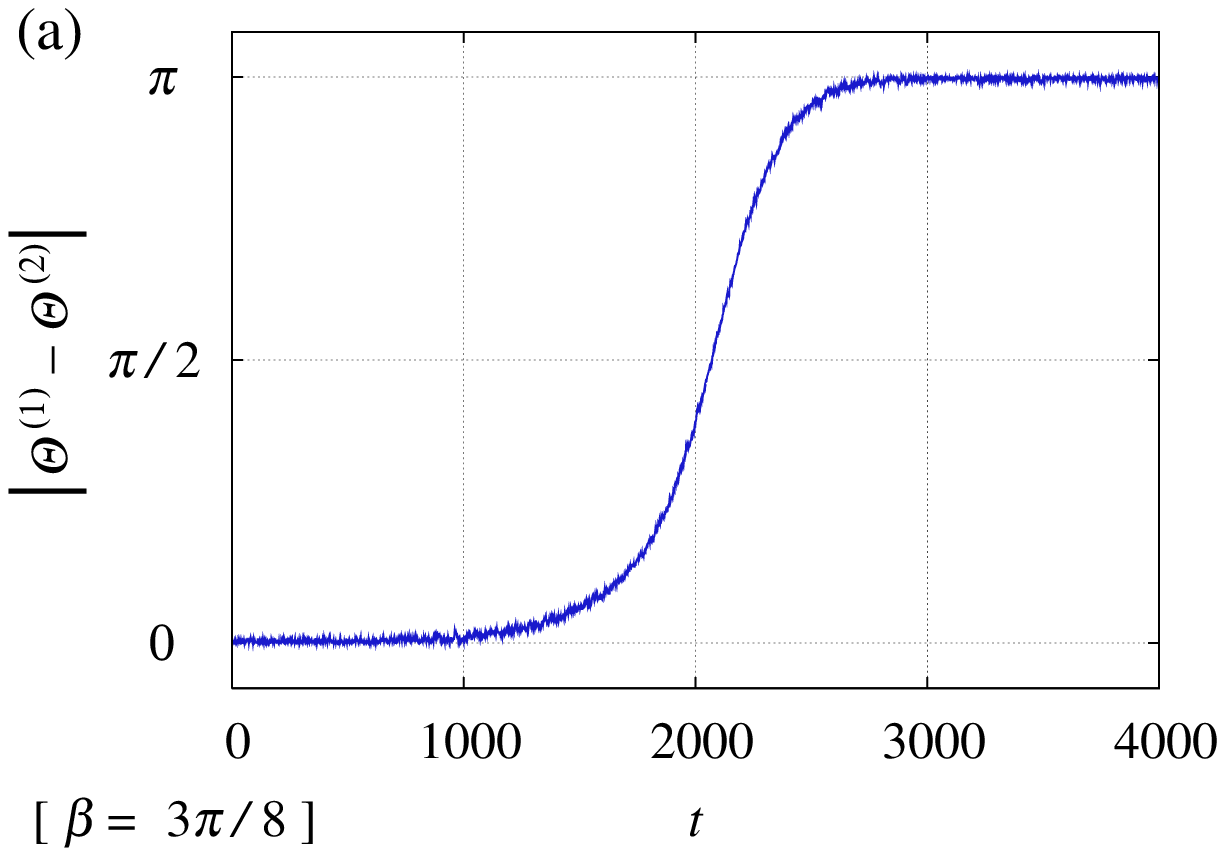}
    \includegraphics[width=0.45\hsize,clip]{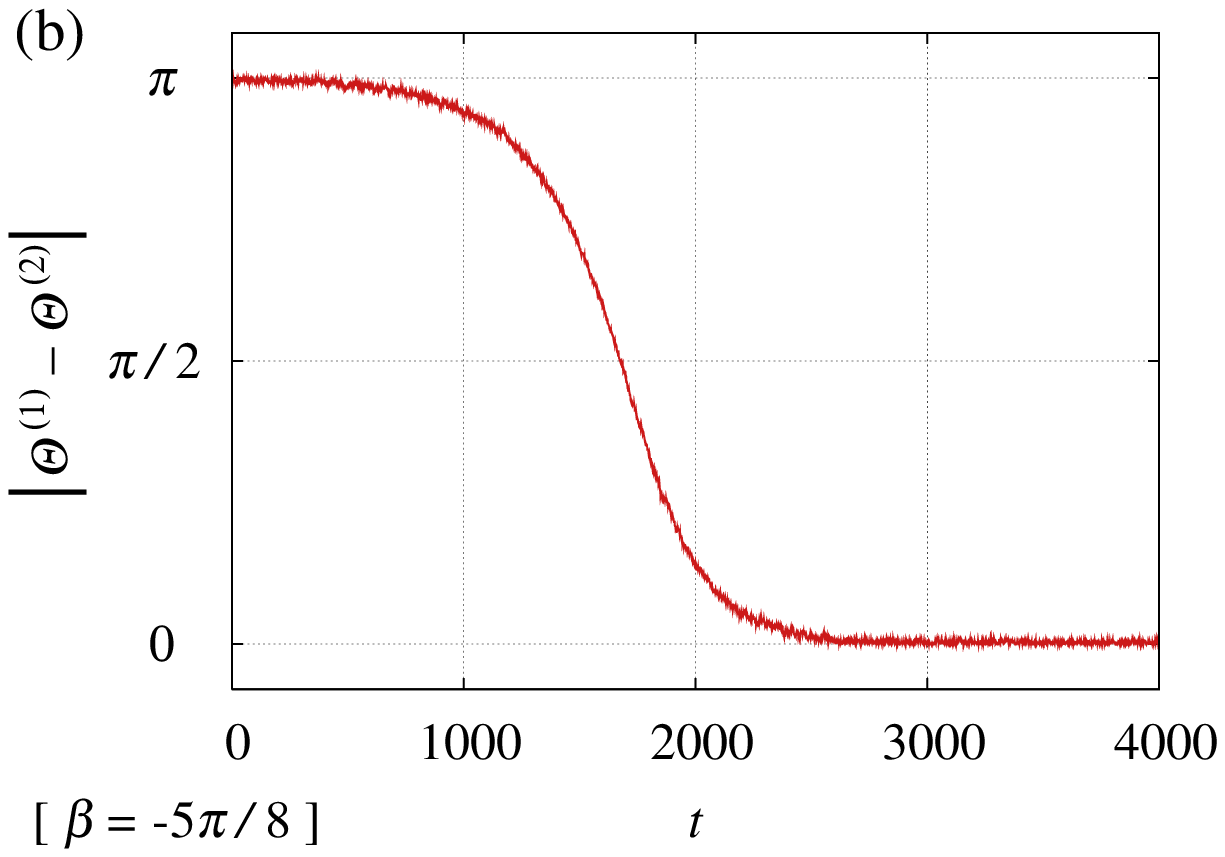}
    \caption{(Color online)
      Time evolution of collective phase difference
      $| \Theta^{(1)} - \Theta^{(2)} |$.
      (a) Effective anti-phase collective synchronization
      with microscopic in-phase external coupling, $\beta = 3 \pi / 8$.
      (b) Effective in-phase collective synchronization
      with microscopic anti-phase external coupling, $\beta = -5 \pi / 8$.
      The other parameters are $K = J = 1$, $\epsilon = 0.01$,
      $\omega_0 = 0$, $\gamma = ( K \cos \alpha ) / 4$,
      and $\alpha = 3 \pi / 8$.
      The number of oscillators in each group is $N = 4096$.
    }
    \label{fig:1}
  \end{center}
\end{figure}

\begin{figure}
  \begin{center}
    \includegraphics[width=0.45\hsize,clip]{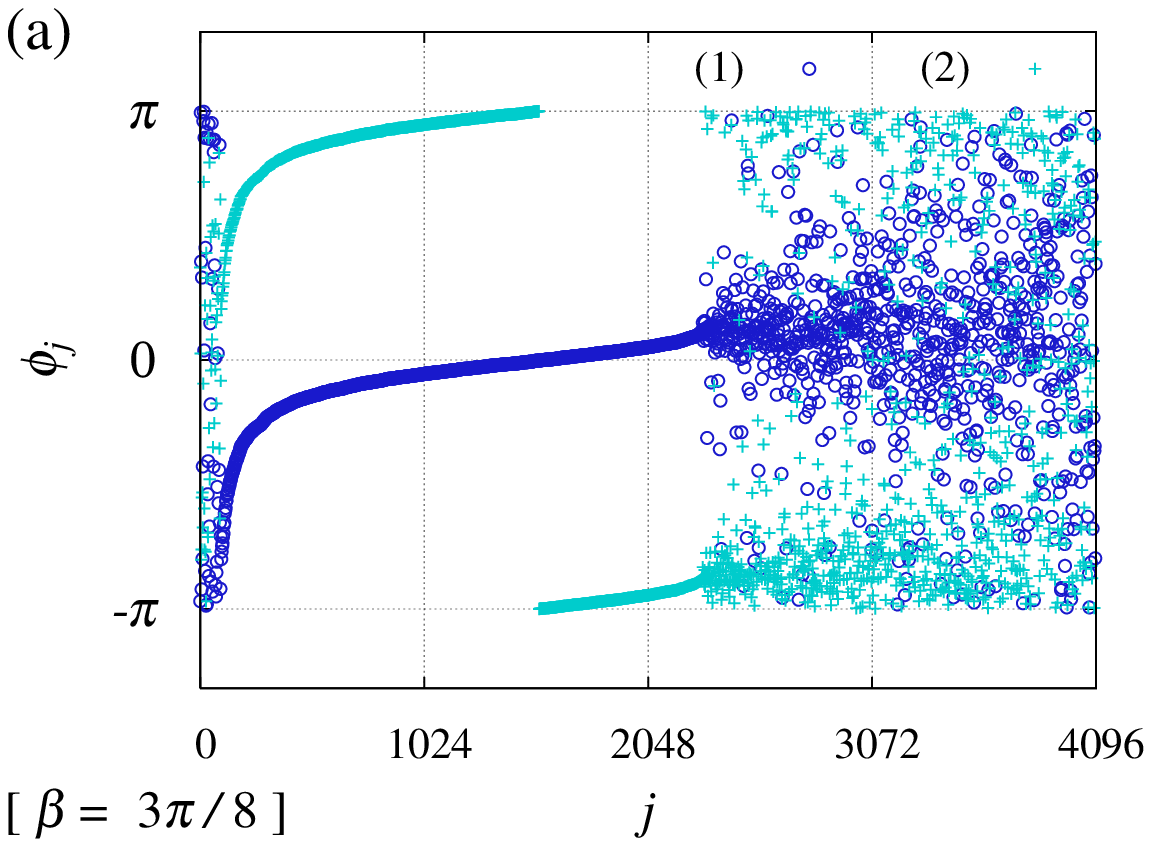}
    \includegraphics[width=0.45\hsize,clip]{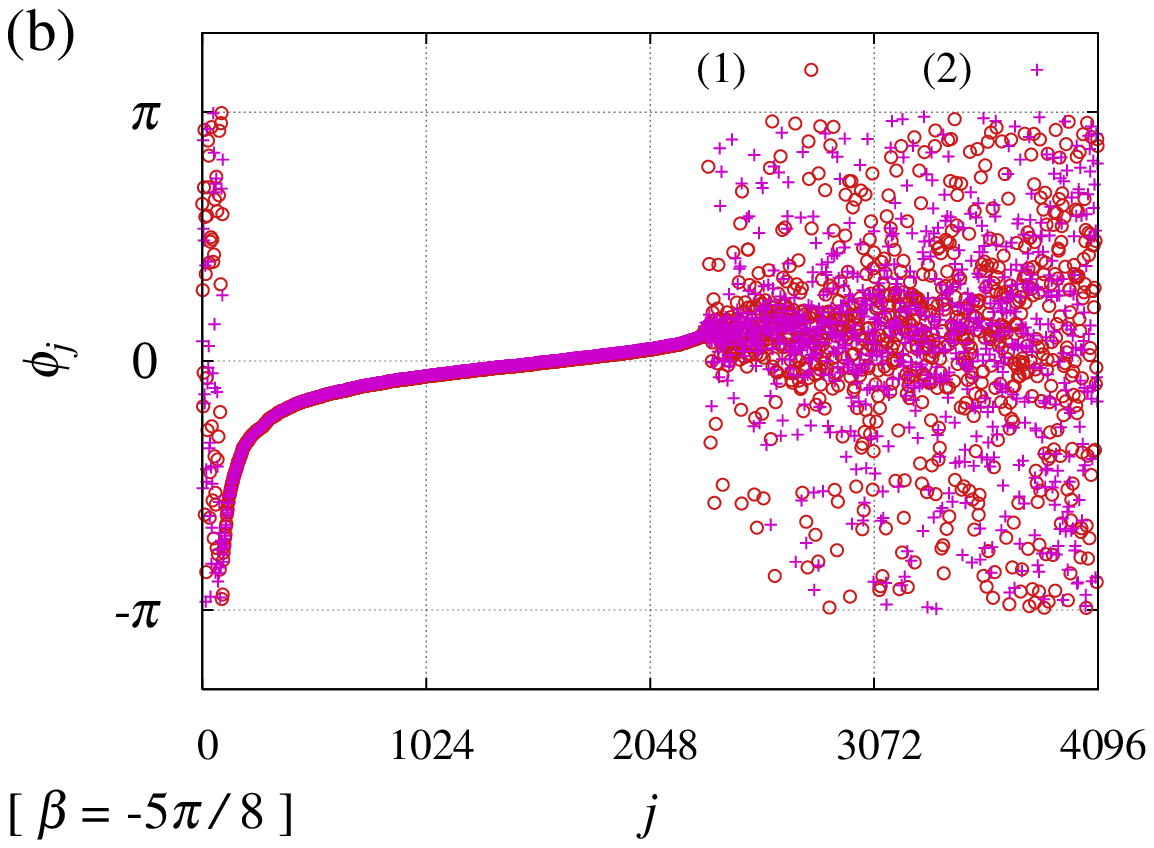}
    \caption{(Color online)
      Snapshots of the asymptotic states of the oscillators in Fig.~\ref{fig:1}.
      The oscillators in each group are sorted in increasing order of their natural frequencies.
      Only one in every two oscillators is plotted.
      Open circles ($\circ$) and plus signs ($+$) indicate
      oscillator in group~$(1)$ and in group~$(2)$, respectively.
      The collective frequency of each group is $\Omega = (-3/4)\sin(3\pi/8)$,
      which differs from the central frequency $\omega_0 = 0$.
      (a) Effective anti-phase coupling with microscopic in-phase coupling, $\beta = 3 \pi / 8$.
      (b) Effective in-phase coupling with microscopic anti-phase coupling, $\beta = -5 \pi / 8$.
    }
    \label{fig:2}
  \end{center}
\end{figure}

\begin{figure}
  \begin{center}
    \includegraphics[width=0.55\hsize,clip]{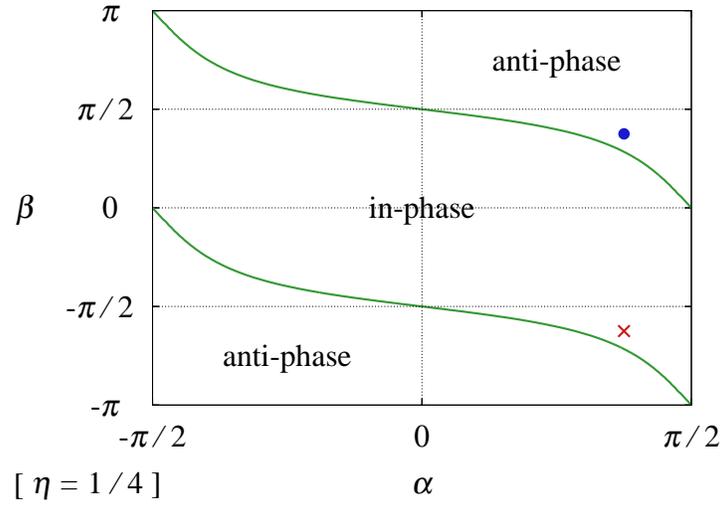}
    \caption{(Color online)
      A diagram showing whether the collective phase coupling function between the groups
      is in-phase or anti-phase on the $\alpha$-$\beta$ plane for
      $\alpha \in ( -\pi / 2, \pi / 2 )$, $\beta \in [ -\pi, \pi ]$, and $\eta = 1 / 4$.
      The solid curves are theoretically determined from
      Eq.~(\ref{eq:type2}), i.e., by $\rho \cos \delta = 0$.
      The filled circle ($\bullet$) indicates
      $\alpha = \beta = 3 \pi / 8$
      corresponding to Figs.~\ref{fig:1}(a) and \ref{fig:2}(a),
      and the cross ($\times$) indicates
      $\alpha = 3 \pi / 8$ and $\beta = -5 \pi / 8$
      corresponding to Figs.~\ref{fig:1}(b) and \ref{fig:2}(b).
    }
    \label{fig:3}
  \end{center}
\end{figure}

Now, let us reexamine the case with $\eta = 1/4$,
which we used in the numerical simulations displayed in Fig.~\ref{fig:1}.
The types of the collective phase coupling function of Eq.~(\ref{eq:digamma2})
is shown in Fig.~\ref{fig:3} on the $\alpha$-$\beta$ parameter plane,
where the solid curves represent boundaries between the in-phase and anti-phase regimes.
The curves show the marginal condition $\rho \cos \delta = 0$,
which is determined from Eq.~(\ref{eq:type2}).
Two sets of parameter values used in generating Fig.~\ref{fig:1} are plotted in Fig.~\ref{fig:3}.
As can be seen,
the set of parameters corresponding to Fig.~\ref{fig:1}(a) is in the effective anti-phase regime,
whereas that corresponding to Fig.~\ref{fig:1}(b) is in the effective in-phase regime.
Therefore, the theory developed in Sec.~\ref{sec:theory} successfully explains
the numerical results displayed in Fig.~\ref{fig:1}.

\section{Concluding remarks} \label{sec:conclusion}

Three cases of macroscopic phase descriptions for
collective oscillations exhibited by coupled phase oscillator systems have been established for
(i) phase coherent states in nonlocally coupled noisy identical oscillators,
(ii) fully phase-locked states in networks of coupled noiseless non-identical oscillators, and
(iii) partially phase-locked states in globally coupled noiseless non-identical oscillators.
Here, the case~(ii) can be fully analyzed from the viewpoint of dynamical systems,
while the other cases (i) and (iii) necessitate statistical treatments.

The collective phase dynamics of the case~(i) was established
in Refs.~\cite{ref:kawamura07,ref:kawamura08,ref:kawamura10},
where the collective phase equation was derived for the first time.
In this case, it is essential to derive a nonlinear Fokker-Planck equation
from coupled Langevin phase equations by using the mean-field theory~\cite{ref:kuramoto84},
which is applicable for nonlocal coupling as well as for global coupling
in a large population of identical oscillators with independent noise.
Applying the phase reduction method to the nonlinear Fokker-Planck equation,
we can derive the collective phase equation.
Furthermore, using the center-manifold reduction method in addition to the phase reduction method,
a detailed analysis can be performed near the onset of collective oscillations via the supercritical Hopf bifurcation.

The collective phase description for the case~(ii) was formulated in Ref.~\cite{ref:kori09}.
In this case, we can systematically treat
any system size, connectivity, heterogeneity in the coupling, and nonuniform external forcing,
as long as the oscillators exhibit fully phase-locked collective oscillations.
In particular, the Jacobi matrix of the collectively oscillating solution
takes the form of the Laplacian matrix encountered in graph theory~\cite{ref:biggs97,ref:agaev00},
so that several analytical results can be obtained by using the matrix tree theorem~\cite{ref:masuda09a,ref:masuda09b}.
There exist several studies related to this case~\cite{ref:ko09,ref:toenjes09}
(see also Refs.~\cite{ref:ermentrout92,ref:mirollo05}).

The present paper provides a tractable example of the collective phase description for the case~(iii).
The keystone in our analysis is the Ott-Antonsen ansatz~\cite{ref:ott08,ref:ott09},
which is unfortunately limited to the case with global sinusoidal coupling and Lorentzian frequency distributions,
but which yields analytically tractable coupled Stuart-Landau equations for the complex order parameters.
By virtue of the circular symmetry of the limit-cycle,
we could explicitly calculate the collective phase coupling function between the groups.

However, a general framework for collective phase reduction in the case~(iii) is still missing.
It would be necessary to derive a continuity equation, such as the nonlinear Fokker-Planck equation,
similarly to the case~(i), which can easily be done.
However, the fundamental difficulty in applying the phase reduction method to the continuity equation in this case
lies in the fact that the zero eigenvalue corresponding to the collective phase mode
may not be isolated, but immersed in the continuous spectrum on the imaginary axis,
as implied by the linear stability analysis~\cite{ref:mirollo07}.
In the formulations of the cases (i) and (ii),
it is critically important that the zero eigenvalue corresponding to the collective phase mode is isolated.
In the present study,
reduction of the infinite-dimensional phase space to a finite-dimensional manifold by using the Ott-Antonsen ansatz
yielded an isolated zero eigenvalue corresponding to the collective phase mode.
But it is an open problem at this point how to extend the present analysis
to more general phase coupling functions with higher harmonic terms
and to more general frequency distributions of the oscillators.

In conclusion, we have established an analytically tractable example
of the collective phase description of globally coupled non-identical phase oscillators.
We have found that the type of the collective phase coupling function can be different from
that of microscopic external coupling function,
and clarified the relation between them by systematically deriving the collective phase equation
from the microscopic phase equations.
The collective phase reduction would serve as a powerful method
in analyzing meta-groups of coupled oscillators comprised of multiple interacting groups.

\appendix

\section{Self-consistent theory of collective oscillations} \label{sec:A}

As a validation of our arguments based on the Ott-Antonsen ansatz~\cite{ref:ott08,ref:ott09},
we compare the limit-cycle solution given in Eq.~(\ref{eq:solution})
with the result obtained by a conventional self-consistent theory~\cite{ref:kuramoto84}.
As is well known, the self-consistent equation~\cite{ref:kuramoto75,ref:kuramoto84,ref:sakaguchi86}
for the order parameter amplitude $R$ of Eq.~(\ref{eq:simulation}) with $\epsilon = 0$,
is given in the following form (see also Refs.~\cite{ref:kuramoto02,ref:shima04}):
\begin{equation}
  R e^{i\alpha} = \int_{-\infty}^{\infty} d\omega \, g(\omega)
  \left[ \sqrt{1 - \left( \frac{\omega - \Omega}{KR} \right)^2}
    + i \left( \frac{\omega - \Omega}{KR} \right) \right],
  \label{eq:self-consistent}
\end{equation}
where contributions from both coherent and incoherent parts are expressed
in a single formula~\cite{ref:abrams04,ref:abrams06,ref:tsubo07}.
There is a unique eigenvalue $\Omega$ of the collective frequency
for which the self-consistent equation~(\ref{eq:self-consistent})
of the order parameter amplitude $R$ admits a solution.
For the Lorentzian distribution of Eq.~(\ref{eq:lorentzian}),
we can analytically solve the self-consistent equation~(\ref{eq:self-consistent}) as follows.
No such calculation seems to have been carried out so far.

The Lorentzian distribution of Eq.~(\ref{eq:lorentzian}) can be expressed by
\begin{equation}
  g\left( \omega \right) = \frac{\gamma}{\pi}
  \frac{1}{[ ( \omega - \omega_0 ) + i \gamma ][ ( \omega - \omega_0 ) - i \gamma ]}.
\end{equation}
Taking the upper half-plane as the contour of integration for Eq.~(\ref{eq:self-consistent}),
we obtain
\begin{equation}
  R e^{i\alpha} = \sqrt{1 - z^2} + i z, \qquad
  z \equiv \frac{(\omega_0 - \Omega) + i \gamma}{KR}.
\end{equation}
This equation can be transformed into
\begin{equation}
  R^2 e^{i\alpha} - 2iRz - e^{-i\alpha} = 0,
\end{equation}
which is equivalent to the following simultaneous equations:
\begin{align}
  R^2 \cos\alpha + \frac{2\gamma}{K} - \cos\alpha &= 0, \\
  R^2 \sin\alpha - \frac{2(\omega_0 - \Omega)}{K} + \sin\alpha &= 0.
\end{align}
Solving these equations, the amplitude and frequency of the collective oscillation
can be respectively obtained as
\begin{align}
  R^2 &= 1 - \frac{2\gamma}{K\cos\alpha}, \\
  \Omega &= \omega_0 - K \sin\alpha + \gamma \tan\alpha,
\end{align}
which coincide with the results given in Eq.~(\ref{eq:solution}).

\section{Derivation of the collective phase sensitivity function} \label{sec:B}

We consider a group of globally coupled non-identical phase oscillators
subject to common weak external forcing, $\epsilon p(t)$,
described by the following equation:
\begin{equation}
  \dot{\phi}_j = \omega_j - \frac{K}{N}
  \sum_{k=1}^N \sin\left( \phi_j - \phi_k + \alpha \right)
  + \epsilon Z\left( \phi_j \right) p(t),
\end{equation}
where the microscopic phase sensitivity function~\cite{ref:kuramoto84} is taken as
\begin{equation}
  Z\left( \phi \right) = - \sin\phi.
\end{equation}
The Ott-Antonsen ansatz is applicable also in this global sinusoidal case~\cite{ref:ott08,ref:ott09}.
Therefore, we can derive the amplitude equation for the complex order parameter $A(t)$
in the following form:
\begin{equation}
  \dot{A} = \left( \mu + i \Omega_{\rm c} \right) A - g \left| A \right|^2 A
  + \epsilon \frac{1}{2} \left( 1 - A^2 \right) p(t).
\end{equation}
By applying the phase reduction method, the collective phase equation is obtained as
\begin{equation}
  \dot{\Theta} = \Omega + \epsilon \zeta( \Theta ) p(t),
\end{equation}
where we assumed that the external forcing is sufficiently weak.
The collective phase sensitivity function is given by
\begin{align}
  \zeta\left( \Theta \right)
  &= {\rm Re}\, \left[ \bar{U}_0^\ast\left( \Theta \right)
    \frac{1}{2} \left\{ 1 - \bigl( A_0\left( \Theta \right) \bigr)^2 \right\} \right] \nonumber \\
  &= -\left( \frac{R_0+R_0^{-1}}{2} \right) \sin\Theta
  + \frac{{\rm Im}\, g}{{\rm Re}\, g} \left( \frac{R_0-R_0^{-1}}{2} \right) \cos\Theta,
\end{align}
which is sinusoidal in form.
In general, the collective phase sensitivity function $\zeta(\Theta)$
differs from the microscopic phase sensitivity function $Z(\phi)$.
When the oscillators become completely phase synchronized, $\eta = 0$, i.e., $R_0 = 1$,
the collective phase sensitivity function coincides with the microscopic one,
$\zeta(\Theta) = Z(\Theta)$.
Near the onset of collective oscillations, $\eta \lesssim 1/2$, i.e., $\mu \gtrsim 0$,
the amplitude of the collective phase sensitivity function increases as
$\zeta(\Theta) = O(\mu^{-1/2})$.



\clearpage


\end{document}